\documentclass[12pt,amsart,superscriptaddress]{iopart}
\usepackage{fullpage,graphicx,epsf}
\usepackage{iopams}%
\usepackage{setstack}
\usepackage{amsfonts,amssymb}
\usepackage{amscd}
\usepackage{cite}
\usepackage{color}%
\usepackage{graphicx}
\usepackage{epsfig}
\usepackage[english]{babel}
\usepackage{amsfonts}
\usepackage{soul}
\begin{document}

\def\l{\label}
\def\p{\partial}

%=============================================================================
\title{Kinetic theory for non-equilibrium stationary states in long-range
interacting systems}
\author{Cesare Nardini$^{1,2}$, Shamik Gupta$^1$, Stefano Ruffo$^{1,3}$,
Thierry Dauxois$^{1}$ and Freddy Bouchet$^1$}
\address{$^1$ Laboratoire de Physique de l'Ecole Normale Sup\'erieure de Lyon, Universit\'e de Lyon and CNRS, 46, all\'ee d'Italie, F-69007 Lyon, France}
\address{$^2$ Dipartimento di Fisica e Astronomia and CSDC, Universit\`a di Firenze and INFN, via G. Sansone 1, I-50019 Sesto Fiorentino (FI), Italy}
\address{$^3$ Dipartimento di Energetica ``Sergio Stecco" and CSDC,
Universit\`a di Firenze, CNISM and INFN, via S. Marta 3, 50139 Firenze, Italy}
\ead{cesare.nardini@gmail.com,shamik.gupta@ens-lyon.fr,\\stefano.ruffo@gmail.com,thierry.dauxois@ens-lyon.fr,freddy.bouchet@ens-lyon.fr}

\begin{abstract}
We study long-range interacting systems perturbed by external stochastic
forces. Unlike the case of short-range systems, where stochastic forces
usually act locally on each particle, here we consider perturbations by
external stochastic fields. The system reaches stationary states where external forces balance dissipation on average. These states do not respect detailed balance and support non-vanishing fluxes of conserved quantities. 
We generalize the kinetic theory of isolated long-range systems to
describe the dynamics of this non-equilibrium problem. The kinetic
equation that we obtain applies to plasmas, self-gravitating systems, and
to a broad class of other systems. Our theoretical results hold for
homogeneous states, but may also be generalized to apply to inhomogeneous states.
We obtain an excellent agreement between our theoretical predictions and numerical simulations. We discuss possible applications to describe non-equilibrium phase transitions. 
\end{abstract}
\date{\today}
\maketitle

\section{Introduction}
Most physical systems are out of equilibrium either because of coupling to
thermal baths at different temperatures or because of external
forces that break detailed balance. Studying non-equilibrium stationary
states is an active area of research in modern statistical mechanics. It is indeed a lasting challenge to achieve for
non-equilibrium systems a level of theoretical understanding similar to 
the one established for equilibrium systems~\cite{Derrida:2007,Jarzynski:2008,Dhar:2008}.

In this Letter, we consider systems of particles interacting through
two-body non-integrable potentials, also called long-range
interactions. Examples include plasmas and self-gravitating systems (globular clusters, galaxies), where particles
interact through repulsive or attractive Coulomb and attractive Newton
potential, respectively. Our
work also applies to a large class of 
models with non-integrable interactions, such as spins, vortices in two
dimensions, and others, which have been studied extensively in
recent years~\cite{Campa:2009,Bouchet:2010,Chavanis,JStatMech,Bouchet-Venaille}.

Systems with non-integrable potentials are often forced through external
stochastic fields. For example, globular clusters are affected by the
gravitational potential of their galaxy, thereby producing a force that
fluctuates along their physical trajectories.
In addition, galaxies themselves feel the random potential of other
surrounding galaxies, and their halos are subjected to transient and periodic perturbations due, for example, to the passing of dwarfs or to orbital decaying~\cite{Weinberg2000}. Plasmas may also be subjected to fluctuating
interactions imposed by environmental electric or magnetic fields~\cite{Liewer}. These
physical situations often lead to a stationary state where
the power injected by the external random fields balances on average the
dissipation. To the best of our knowledge, such non-equilibrium stationary
states in systems with non-integrable
potentials have not been studied before, and this work provides a first step in this direction. 

Unlike systems with short-range interactions, stochastic perturbations
in long-range interacting systems often act coherently on all particles  and not
independently on each particle. Moreover, unlike short-range systems, it
is not natural to consider long-range systems as being
coupled to thermal baths at the boundaries. Thus, the non-equilibrium
stationary states that we study are rather different from the ones in
systems with short-range interactions. These states
do not verify detailed balance and support non-zero fluxes of
conserved quantities, which are basic ingredients of non-equilibrium stationary states. 

Theoretical results on isolated systems with
long-range interactions include the kinetic theory description of 
relaxation towards equilibrium. In plasma physics, this approach leads
to the Lenard-Balescu equation or to the approximate Landau equation~\cite{Nicholson:1992,Lifshitz:2002}.
These equations, or some of their approximations, are grouped as the
collisional Boltzmann equation in the astrophysical context. 
The main theoretical result of this Letter is a generalization of the kinetic
theory to describe non-equilibrium stationary states, valid for small
external perturbations and spatially homogeneous stationary states. 

The non-equilibrium kinetic equation that we obtain describes the temporal evolution of
the one-particle distribution function. When the system is not far from
equilibrium, it is natural to expect that the system settles into a
stationary state. We find that in such a state, the one-particle
momentum distribution is non-Gaussian. The kinetic equation describes the evolution
of the kinetic energy, and its prediction of the stationary state compares very well with $N$-body numerical simulations.  

%=============================================================================
\section{Stochastically forced long-range interacting systems}
Consider a system of $N$ particles interacting through a long-ranged
pair potential. The Hamiltonian of the system is
\begin{equation}\l{Hamiltonian}
H=\sum_{i=1}^N\frac{p_i ^2}{2}+\frac{1}{2N}\sum_{i,j=1}^N v(q_i-q_j),
\end{equation}
where $q_i$ and $p_i$ are, respectively, the coordinate and the momentum
of the $i$-th particle, while $v(q)$ is the two-body interaction
potential. The particles are taken to be of unit mass. In this paper, for simplicity, we consider $q_i$'s to be
scalar periodic variables of period $2\pi$; generalization to $q_i \in
\mathbb{R}^n$, with $n=1, 2$ or $3$, is straightforward.  

In self-gravitating systems, since the dynamics is dominated by
collective effects, it is natural and usual to rescale time in such a
way that the parameter $1/N$ multiplies the interaction potential~\cite{Milion-body-pb}.  In plasma physics, the typical  number of
particles with which one particle interacts is given by the coupling
parameter $\Gamma = n\lambda_D^3$, where $n$ is the number density and
$\lambda_D$ is the Debye length. It is then usual to rescale the time
such that the inverse of a power of $\Gamma$ multiplies the interaction
term~\cite{Nicholson:1992}. These reasons justify the rescaling of the
potential energy by $1/N$ in Eq.~(\ref{Hamiltonian}), known as the
Kac scaling in systems with long-range interactions~\cite{Kac}. 

We perturb the system~(\ref{Hamiltonian}) by the action of the stochastic field $F(q_i,t)$. The resulting equations of motion  are
\begin{eqnarray}
\label{equations_of_motion}
\dot{q}_i=\frac{\p H}{\p p_i},\qquad {\rm and} \qquad \dot{p}_i=-\frac{\p H}{\p q_i}-\alpha p_i +\sqrt{\alpha}\,F(q_i,t),
\end{eqnarray}
where $\alpha$ is the friction constant, and $F(q,t)$ is a statistically homogeneous Gaussian
process with zero mean and variance given by 
\begin{equation}
\langle F(q,t)F(q',t')\rangle = C(|q-q'|)\delta(t-t').
\end{equation}
The hypothesis that the Gaussian fields are statistically homogeneous,
i.e., the correlation function depends solely on $|q-q'|$, holds
for any perturbation which does not break space homogeneity. Such a
hypothesis will
also be essential for the following discussions where we consider homogeneous stationary states. Now, $C(q)$ represents
correlation, and is therefore a positive-definite function~\cite{Papoulis}.
Its Fourier components are thus positive:
\begin{equation}\label{eq:g}
c_k \equiv \frac{1}{2\pi}\int_0^{2\pi}{\rm d}q ~C(q)e^{-ikq} > 0,\ \qquad
\qquad C(q)=c_0+2\sum_{k=1}^{\infty}c_k \cos(kq).
\end{equation}
It will be convenient to use the following equivalent Fourier representation of
the Gaussian field $F(q,t)$:
\begin{eqnarray}\label{stoc_forces} 
F(q,t)\,=\,\sqrt{c_0}\ X_0 + \sum_{k=1}^{\infty} \sqrt{2 c_k}\left[\cos (kq) \,X_k + \sin
(kq) \, Y_k\right],
\end{eqnarray}
where $X_k$ and $Y_k$ are independent scalar Gaussian white noises satisfying
$\langle X_k(t)\,X_{k'}(t')\rangle=\delta_{k,k'}\delta(t-t')$,
$\langle Y_k(t)\,Y_{k'}(t')\rangle=\delta_{k,k'}\delta(t-t')$, and
$\langle X_k(t)\,Y_{k'}(t')\rangle=0$.

Using the It\={o} formula~\cite{Gardiner} to compute the time derivative of
the energy density $e=H/N$ and averaging over noise realizations
give
\begin{equation}\label{evolution_H}\label{kinetic-energy} 
\left\langle\frac{de}{dt}\right\rangle+\left \langle 2\alpha \kappa\right\rangle=\frac{\alpha}{2}C(0)\,,
\end{equation}
where $\kappa=\sum_{i=1}^Np_i ^2/(2N)$ is the kinetic energy density. The
average kinetic energy density in the stationary state is thus
$\left\langle \kappa\right\rangle_{ss} = C(0)/4$.

In the dynamics~(\ref{equations_of_motion}), fluctuations of intensive observables due
to stochastic forcing are of order $\sqrt{\alpha}$, while those due to
finite-size effects are of order $1/\sqrt{N}$. Moreover, the typical timescale
associated with the effect of stochastic forces is $1/\alpha$ (see,
e.g., Eq.~(\ref{kinetic-energy})), while the one associated with
relaxation to equilibrium due to finite-size effects is of order $N$,
see~\cite{Campa:2009,Bouchet:2010}.   

In the following, we analyze the dynamics 
(\ref{equations_of_motion}) in the joint limit
$N\to \infty$ and $\alpha \to 0$. While the first limit is physically
motivated on grounds that most long-range systems indeed contain a large
number of particles, the second one allows us to study non-equilibrium
stationary states for small external forcing. Moreover, for small
$\alpha$, we will be able to develop a complete kinetic theory for the dynamics.

For simplicity, we discuss in this Letter the continuum limit  $N\alpha
\gg 1$, when stochastic effects are predominant with respect to
finite-size effects. Generalization to other cases ($N\alpha$ of order
one, or, $N\alpha \ll 1$) is straightforward, as discussed in the conclusion.

\section{Kinetic theory}
\label{kinetic_theory}

A natural framework to study the dynamics~(\ref{equations_of_motion}) is the
kinetic theory. We now describe the theoretical approach to derive this
theory, while some of the technical results will be explicitly obtained in a longer paper~\cite{long-version}. The central result is the kinetic equation
(\ref{kinetic_equation}) below, which describes the time evolution of the single-particle distribution function. 

We consider the Fokker-Planck equation associated with the equations of motion
(\ref{equations_of_motion}). It describes the evolution of the
$N$-particle distribution function $f_N(q_1,...,q_N,p_1,...,p_N,t)$ (after averaging over the noise
realization, $f_N$ is the probability density
to observe the system with coordinates and momenta around the values
$\{q_i,p_i\}_{1 \leq i \leq N}$ at time $t$). This equation can be
derived by standard methods~\cite{Gardiner}.  We get 
\begin{eqnarray}
\l{F-P-eq}
&&\frac{\p f_N}{\p t}=\sum_{i=1}^N\left[-p_i\frac{\p f_N}{\p q_i}+\frac{\p
(\alpha p_i f_N)}{\p p_i}\right]\nonumber \\
&&+\frac{1}{2N}\sum_{i,j=1}^N\frac{\p
v(q_i-q_j)}{\p q_i}\left[\frac{\p}{\p p_i}-\frac{\p }{\p
p_j}\right]f_N+\frac{\alpha}{2}\sum_{i,j=1}^N C(q_i-q_j)\frac{\p ^2
f_N}{\p p_i \p p_j}. 
\end{eqnarray}
We have proved by analyzing the so-called potential
conditions~\cite{Risken} for this Fokker-Planck equation that a
sufficient condition for the stochastic
process~(\ref{equations_of_motion}) to verify detailed balance is that
the Gaussian noise is white in space, that is, $c_k=c$ for all $k$. This condition is not satisfied for a generic correlation function $C$. Steady states  are then true non-equilibrium ones, with non-vanishing currents and a balance between external forces and dissipation.

Similar to the Liouville equation for Hamiltonian systems, the
$N$-particle Fokker-Planck equation is a very detailed description of the
system.  Using kinetic theory, we want to describe the evolution of the one-particle
distribution function $f(z_1,t)=\int \prod_{i=2}^N\,{\rm d}z_i
~f_N(z_1,...,z_N,t)$ (we use the notation
$z_i=(q_i,p_i)$ whenever convenient). Note that the
normalization is $\int {\rm d} z ~f(z,t)=1$. 

In plasmas and self-gravitating systems, due to the long-range nature of
the interactions, the one-particle
distribution function is not affected by the two-particle distribution
function at leading order in $1/N$, and therefore, its evolution is described at leading order by the
Vlasov equation. Finite-size effects however induce weak correlations
whose effects on the long-time evolution of the one-particle distribution 
can be computed self-consistently in the framework of the kinetic theory by using perturbation theory. A complete
treatment of the problem leads to the Lenard-Balescu
equation~\cite{Nicholson:1992,Lifshitz:2002}. In a similar way, for our problem, the evolution will be described at leading order by the
Vlasov equations due to the long-range nature of the interactions. Weak
stochastic forces lead to weak correlations that affect the long-time
evolution. This case can be treated by following a generalized kinetic approach, as we now describe. 

Substituting in the $N$-particle Fokker-Planck equation~(\ref{F-P-eq}) the reduced distribution
function $f_s(z_1,...,z_s,t)=\int \prod_{i=s+1}^N\,{\rm
d}z_i\,f_N(z_1,...,z_N,t)$, and using standard techniques~\cite{Huang}, we get a hierarchy of equations, similar to the BBGKY hierarchy.  We split the reduced
distribution functions into connected and non-connected parts, e.g.,
$f_2(z_1,z_2,t)=f(z_1,t)f(z_2,t)+ \alpha g(z_1,z_2,t)$, and then neglect
the effect of the connected part of the three-particle correlation on
the evolution of the two-particle correlation function. This scheme is
consistent at leading order in the small parameter $\alpha$, and is  the
simplest closure scheme for the hierarchy. For simplicity, we moreover
assume that the system is homogeneous: $f$ depends on $p$, and $g$
depends on $|q_1-q_2|$, $p_1$ and $p_2$, only. The first two equations of the hierarchy are then
\begin{eqnarray}\label{1bbgky_2}
\frac{\p f}{\p t}-\alpha\frac{\p}{\p p}[ p
f]-\frac{\alpha}{2}C(0)\frac{\p^2 f}{\p p^2}= \alpha \frac{\p}{\p p}
\int {\rm d}q {\rm d}p_2 ~v'(q)g(q,p,p_2,t),
\end{eqnarray}
and
\begin{eqnarray}\label{2bbgky_2}
&&\frac{\p g}{\p t}+\left[p_1\frac{\p g}{\p
q_1}-\left.\frac{\p f}{\p p}\right|_{p_1}\int
 {\rm d}q_3 {\rm d}p_3 ~v'(q_1-q_3)g(q_3-q_2,p_3,p_2,t)\right]+\{1 \leftrightarrow
2\}\nonumber \\
&&=C(q_1-q_2)\left.\frac{\p f}{\p p}\right|_{p_1}\left.\frac{\p f}{\p
p}\right|_{p_2},
\end{eqnarray}
where the symbol $\{1 \leftrightarrow 2\}$ means an expression obtained
from the bracketed one by exchanging $1$ and $2$, while the prime denotes
differentiation.

To obtain from these equations a single kinetic equation for the distribution function $f$, we have to solve Eq.~(\ref{2bbgky_2}) for $g$ as a function of $f$ and plug the result into the right hand side of Eq.~(\ref{1bbgky_2}). From these two equations, we readily see that the two-particle correlation $g$ evolves over a timescale of order one, whereas the
one-particle distribution function $f(p,t)$ evolves over a timescale of
order $1/\alpha$. We use this timescale separation, and compute the
long-time limit of $g$ from Eq.~(\ref{2bbgky_2}) by assuming $f$ to be
constant. This procedure is equivalent to making the Bogoliubov's
hypothesis for deriving the
kinetic theory of isolated long-range systems. For
the timescale separation to be valid, it is also required that the one-particle distribution function $f(p,t)$ is a stable 
solution of the Vlasov equation at all times. 

The solution of equations of the type~(\ref{2bbgky_2}) is quite
technical (see the long appendix in Nicholson's book~\cite{Nicholson:1992}).
Equation~(\ref{2bbgky_2}) differs from the corresponding equation for an
isolated long-range system in that the term on the right hand
side is different in the two cases, and cannot be solved by methods
known in the literature. The main
technical achievement that aided this work is to be able to
solve Eq.~(\ref{2bbgky_2}). In a future paper~\cite{long-version}, we
will give the details on the solving procedure. In brief, the method
relies on making a parallel between the Lyapunov
equations for infinite-dimensional Ornstein-Uhlenbeck processes and their
general solutions, and Eq.~(\ref{2bbgky_2}). Using this method, we get the desired kinetic equation:
\begin{eqnarray}
\label{kinetic_equation}
\frac{\p f}{\p t}-\alpha\frac{\p (p f)}{\p
p}-\alpha\frac{\p}{\p p}\left[D[f]\frac{\p f}{\p
p}\right]=0,
\end{eqnarray}
where 
\begin{eqnarray}
\label{diffusion_coefficient}
D[f](p)=\frac{1}{2}C(0)+2\pi\sum_{k=1}^{\infty}v_{k}c_{k}\int^*{\rm
d}p_1~\left[ \frac{1}{|\epsilon(k,k
p)|^2} + \frac{1}{|\epsilon(k,k p_1)|^2} \right]
\frac{1}{p_1-p}\left.\frac{\p f}{\p p}\right|_{p_1}. \nonumber \\
\end{eqnarray}
Here, $v_{k}$  is the $k$-th Fourier coefficient of the pair potential
$v(q)$, the quantity $c_k$ is defined in Eq.~(\ref{eq:g}), while $\int^*$ indicates the Cauchy
integral, and the dielectric function $\epsilon$ is
\begin{equation}
\label{dielectric_function}
\epsilon(k,\omega)=\lim_{\eta\to 0^+}\left[1-2\pi i v_{k}k \int {\rm
d}p~\frac{1}{-i(\omega+i\eta)+ik p}\frac{\p f}{\p p}\right].
\end{equation} 
\vspace{0.5cm} 
The kinetic equation~(\ref{kinetic_equation}) is the central result of
the Letter. 
  
This kinetic equation has the form of a non-linear Fokker-Planck equation,
since the diffusion coefficient $D[f](p)$ itself is a function of the
unknown distribution function $f$. As Eq.~(\ref{diffusion_coefficient})
shows, this coefficient has two parts, namely, (i) a linear part, $C(0)/2$,
which is due to the mean-field effect of the stochastic forces, and (ii)
a non-linear part due to correlations induced in the system by the
stochastic forces. The contributions of different modes of the
stochastic force are independent of each other, that is, contributions
proportional to $c_k$  do not couple with $v_{k'}$ with $k\neq k'$. 

For consistency,  the prediction of the evolution of the kinetic energy
from the kinetic equation has to agree with Eq.~(\ref{evolution_H}). We
have 
checked this by using Eqs.~(\ref{kinetic_equation})
and~(\ref{diffusion_coefficient}), and proving that the integrals in the non-linear part of the diffusion coefficient give no contribution to the kinetic energy. 

\begin{figure}[h]
\centerline{\includegraphics[width=20cm]{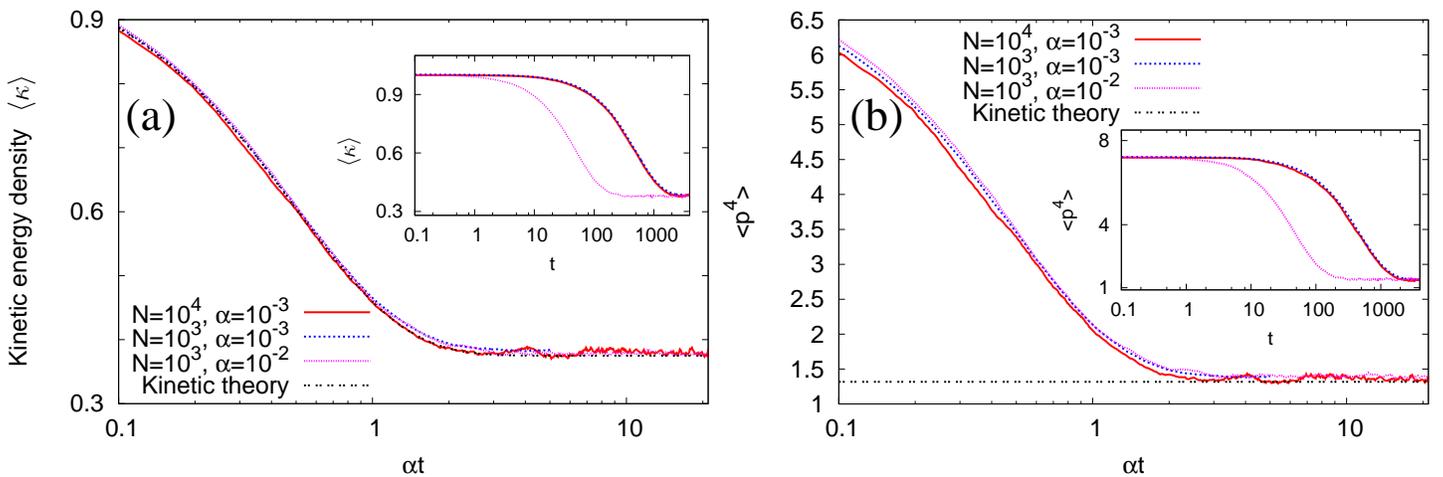}}
\caption{\footnotesize{(a) Kinetic energy density $\langle \kappa
\rangle$ and (b)
$\langle p^4\rangle$ as a function of $\alpha t$, for the values
$C(0)=1.5$ and $c_1=0.75$. The data for different $N$ and $\alpha$
values are obtained from numerical simulations of the stochastically
forced HMF model, and involve averaging over $50$ histories for $N=10^4$
and $10^3$ histories
for $N=10^3$. The data collapse implies that $\alpha$ is the timescale
of relaxation to the stationary state. The inset shows the data without time rescaling by $\alpha$.}}
\label{fig1}
\end{figure}

As already mentioned, and is evident from Eq.~(\ref{kinetic_equation}), the time
scale for the kinetic evolution is $1/\alpha$. This has been checked
by performing direct numerical simulations, see Fig.~\ref{fig1} and the
next section. Thus, $\alpha$ can be eliminated from the kinetic
equation~(\ref{kinetic_equation}) by a redefinition of time. Therefore, even for
vanishingly small value of $\alpha$, if a stationary distribution
exists, it will be at distance of order one from a Gaussian momentum distribution.

While a linear Fokker-Planck equation with non-degenerate diffusion
coefficient can be proven to converge to a unique stationary
distribution~\cite{Risken}, this is not true in general for non-linear
Fokker-Planck equations such as Eq.~(\ref{kinetic_equation}). We expect that if the system is not too far from equilibrium, the kinetic equation will have
a unique stationary state.
Far from equilibrium, the kinetic equation could lead to very
interesting dynamical phenomena, such as bistability, limit cycle or more
complex behaviors. The main issue is then the analysis of the evolution
of the kinetic equation. 
Although some methods to study this type of equations exist
\cite{Frank}, in order to provide some preliminary answers, we have devised a numerical iterative scheme to compute some of the stationary states of the
kinetic equation~(\ref{kinetic_equation}). We now describe the scheme. 

A linear Fokker-Planck equation whose diffusion coefficient $D(p)$ is strictly positive admits a unique stationary state 
\begin{equation}
\label{stationary-FokkerPlanck}
f_{ss}(p)=A\exp\left[-\int_0^p {\rm d}p' ~\frac{p'}{D(p')}\right].
 \end{equation}
For a given distribution $f_n(p)$, we compute the diffusion coefficient
$D_n(p)$ from Eq.~(\ref{diffusion_coefficient}), and then $f_{n+1}$ using $D_n$ and
Eq.~(\ref{stationary-FokkerPlanck}). This procedure defines an iterative scheme. Whenever convergent, this scheme leads to a stationary state of Eq.~(\ref{kinetic_equation}). Each iteration involves integrations, so we expect the method to be robust enough when starting not too far from an actual stationary state. However, we have no detailed mathematical analysis yet. 

In the next section, we discuss numerical results on $N$-particle
simulations, and the computation of stationary states from the
iterative method mentioned above.

\section{Stochastically forced HMF model}
Until now, we have presented our theoretical analysis for a general
two-particle interaction $v(q)$. In order to perform simple numerical simulations, we now consider the case of
the stochastically forced attractive Hamiltonian mean-field (HMF) model, which
corresponds to the choice $v(q)=1-\cos q$. 

The HMF model serves as a paradigm to study long-range
interacting systems, and describes particles moving on a circle under
deterministic Hamiltonian dynamics~\cite{yamaguchi2004,Antoni}. 
This model has been studied a lot in recent times. It displays many features of
generic long-range interacting systems, such as the existence of
quasistationary states~\cite{yamaguchi2004,Campa:2009}. In equilibrium, the
system displays a second-order phase transition from a high-energy homogeneous phase
to a low-energy inhomogeneous phase at the energy density $e_c=3/4$.

Since the Fourier transform of the HMF interparticle potential is, for $k \neq 0$,
$v_k=-\left[\delta_{k,1}+\delta_{k,-1}\right]/2$, where $\delta_{k,i}$
is the Kronecker delta, we see from the kinetic equation
(\ref{kinetic_equation}) that only the stochastic force with wave number
$k=1$ contributes to the non-linear part of the diffusion coefficient; all the other stochastic forces give only a
mean-field contribution through the term
$C(0)$. Thus, the two parameters that dictate the evolution of the
stochastically forced HMF model are $C(0)$ and $c_1$.
From~(\ref{kinetic-energy}), we know that $C(0)=4\langle \kappa
\rangle_{ss}$
is proportional to the kinetic energy in the final stationary state.
Moreover, Eq.~(\ref{eq:g}) implies that $c_1 \le C(0)/2$. 

If $c_1=0$, the kinetic equation reduces to a linear Fokker-Planck
equation with diffusion coefficient $C(0)/2$. This equation also describes the HMF model coupled to a Langevin thermostat,
studied in~\cite{ChavanisBMF,Baldovin_Orlandini}. As the kinetic
equations are the same, the dynamics coincide at leading order in
$\alpha$. However, we know that at higher orders, detailed balance is broken in our case, whereas it holds for the Langevin dynamics. 

In the case $c_1=0$, the homogeneous stationary states of the kinetic
equation have Gaussian momentum distribution $f(p)$.  As has been studied
thoroughly in the context of canonical equilibrium of the HMF model, these states
are stable for kinetic energies greater than $1/4$, i.e., for $C(0)>1$. 

For values of $C(0)$ and $c_1$ such that $C(0)>1$ and $c_1\ll C(0)$,
we then expect the stationary states to be close to homogeneous states
with Gaussian momentum, so that the
numerical iterative scheme to locate stationary states of the kinetic
equation is expected to converge for well-chosen initial conditions. We have checked the convergence for
the set of values of $c_1$ used in
the simulations reported in the paper.  

To check the theory, we have performed numerical simulations of the
stochastically forced HMF model. In Fig. \ref{fig1}, we show the
evolution of the kinetic energy and $\langle p^4
\rangle=(1/N)\sum_{i=1}^N p_i ^4$, and compare them with theoretical
predictions. In the latter case, we have
compared the long-time asymptotic value with the kinetic theory
prediction for the stationary state, computed using the
iterative scheme. In both
cases, we observe a very good agreement between the theory and simulations.

\begin{figure}[here!]
\centering
\begin{tabular}{lr}
\parbox[l]{7.5cm}{
\includegraphics[width=84mm]{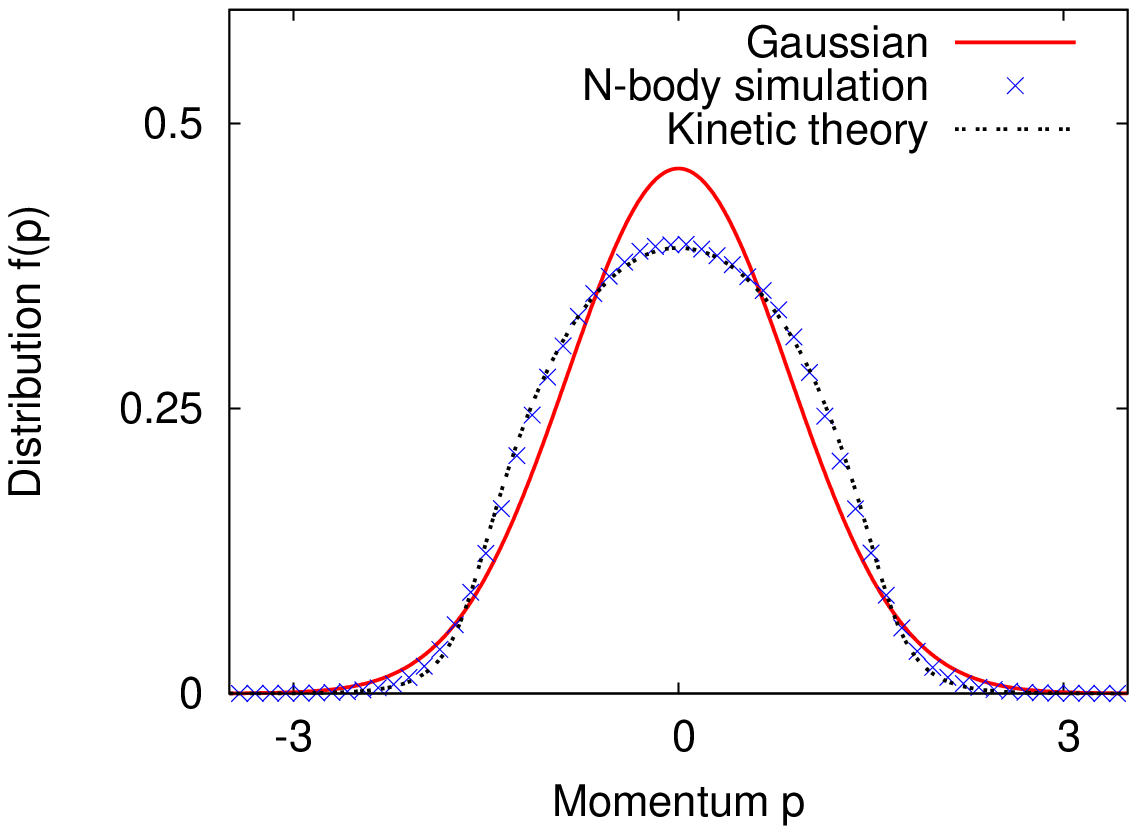}
}&
\parbox[r]{7.5cm}{
\includegraphics[width=84mm]{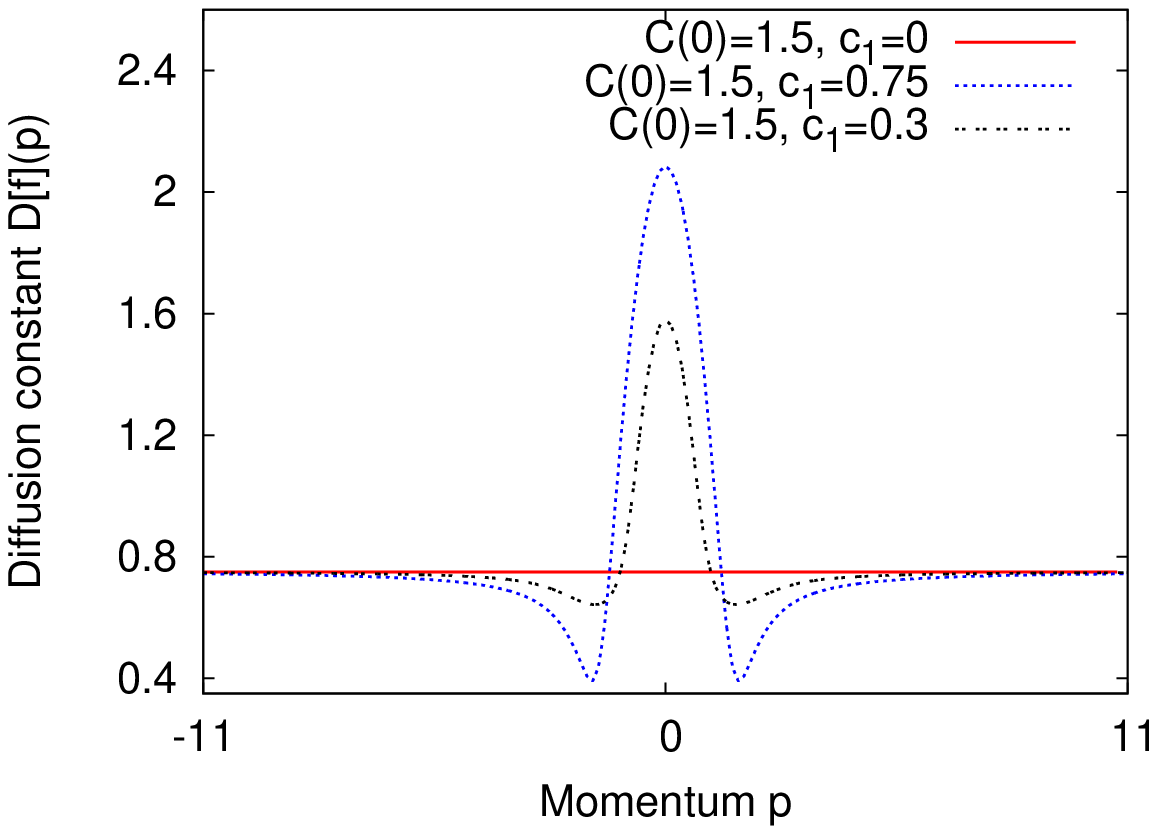}
}
\end{tabular}
\caption{The figure on the left shows the stationary momentum distribution $f(p)$ for $\alpha=0.01$,
$C(0)=1.5$, and $c_1=0.75$. The data denoted by crosses are
results of $N$-body simulations of the stochastically forced HMF model
with $N=10000$, while the black broken line refers to the theoretical
prediction from the kinetic theory. For comparison, the red continuous line shows the Gaussian
distribution with the same kinetic energy (stationary state at
$C(0)=1.5$, $c_1=0$). The figure on the right shows the diffusion
coefficient $D[f](p)$ for the stationary momentum distribution
$f(p)$ for different values of $C(0)$ and $c_1$.}
\label{fig2}
\end{figure}

For a more accurate comparison, we have obtained the stationary momentum
distribution from both $N$-body simulations and the numerical
iterative scheme. The comparison between the two is
shown in Fig. \ref{fig2}, left panel, where we also show the Gaussian distribution
with the same kinetic energy. The agreement between theory and
simulations is excellent.

\section{Conclusions}
In this work, we studied the effect of external stochastic fields on
Hamiltonian long-range interacting systems by generalizing the kinetic
theory of isolated long-range systems. Our theoretical results are general,
being applicable to any long-range inter-particle potential, space dimensions and boundary conditions.
In this paper, we demonstrated an excellent agreement between the theory and numerical
simulations for one representative case. 

Here, we discussed the kinetic theory in the  limit
$N\alpha \gg 1$. The extension to general values of $N\alpha$ is
straightforward:
Because of the linearity of the equations of the BBGKY hierarchy, the
finite-$N$ and stochastic effects give independent contributions. The kinetic equation at leading order of both stochastic and finite-size effects is
\begin{eqnarray}\label{kinetic_equation_general}
\frac{\p f}{\p t} = L_{\alpha}[f] + L_N[f]\,, 
\end{eqnarray}
where $L_{\alpha}$ is the operator described in Eq.
(\ref{kinetic_equation}) and $L_N$ (of order $1/N$) is the
Lenard-Balescu operator~\cite{Nicholson:1992}. For instance, in the case
$N\alpha \ll 1$ and in dimensions greater than one, the operator $L_N$
is responsible for the relaxation to Boltzmann equilibrium after a timescale of order $N$, whereas the smaller effect of $L_{\alpha}$ selects the actual temperature after a longer timescale of order $1/\alpha$.

We note that an equivalent approach to derive the kinetic theory is to
write an evolution equation for the noise-averaged empirical density
$\rho(p,q,t)=(1/N)\sum_{i=1}^N \langle \delta(q_i(t)-q)\delta(p_i(t)-p)
\rangle$, by analogy with the Klimontovich approach for isolated
systems.  The noise appears in the resulting equation as a
multiplicative term. This equation can be treated perturbatively, and
may be shown to lead to the kinetic equation (\ref{kinetic_equation}).  

Let us mention some open issues. For technical simplicity, we assumed a
homogeneous state in our approach. Recently, Heyvaerts~\cite{Heyvaerts}
has generalized the Lenard-Balescu equation to some non-homogeneous
cases; his approach could be used to generalize the theory developed
here to inhomogeneous states. There is no difficulty in principle,
although actual computation could be more involved.

An interesting follow up of this work is to study the dynamics of the kinetic
equation~(\ref{kinetic_equation}), both analytically and numerically.
This may unveil very interesting behaviors, such as bistability or
limit cycles. Bistability was observed in two-dimensional
turbulence with stochastic forcing~\cite{Bouchet-Simonnet}, in a
framework
which has deep connection with the one studied in this Letter. One of
the motivations for this work was to make a first step in formulating a
kinetic theory for the point vortex model and the Euler equations
in two-dimensional turbulence~\cite{Bouchet-Venaille}. This subject will be the topic of further investigations.

%=============================================================================
\section{Acknowledgments}
\vspace{-0.2cm}
C. N. acknowledges the EGIDE scholarship funded by Minist\`ere
des Affaires \'Etrang\`eres. S. G. and S. R. acknowledge
the contract LORIS (ANR-10-CEXC-010-01). F. B. acknowledges the ANR
program STATOCEAN (ANR-09-SYSC-014). Numerical simulations were done at
PSMN, ENS-Lyon. We thank Hugo Touchette for comments on
the manuscript.
\vspace{-0.5cm}
%=============================================================================
\section*{References}
%=============================================================================

%=============================================================================
%=============================================================================
\end{document}